# AB Blanket for Cities*
## (for continual pleasant weather and protection from chemical , biological and radioactive weapons)

## Alexander Bolonkin

C&R, 1310 Avenue R, #F-6, Brooklyn, NY 11229, USA
T/F 718-339-4563, aBolonkin@juno.com, http://Bolonkin.narod.ru

## Abstract

In a series of previous articles (see references) the author offered to cover a city or other important large installations or subregions by a transparent thin film supported by a small additional air overpressure under the form of an AB Dome. That allows keeping the outside atmospheric conditions (for example weather) away from the interior of the inflatable Dome, protecting a city by its' presence from chemical, bacterial, and radioactive weapons and even partially from aviation and nuclear bombs.

The building of a gigantic inflatable AB Dome over an empty flat surface is not difficult. The cover is spread on a flat surface and a ventilator pumps air under the film cover and lifts the new dome into place (inflation takes many hours). However, if we want to cover a city, garden, forest or other obstacle course (as opposed to an empty, mowed field) we cannot easily deploy the thin film over building or trees without risking damage to it by snagging and other complications. In this article is suggested a new method which solves this problem. The idea is to design a double film blanket filled by light gas (for example, methane, hydrogen, or helium - although of these, methane will be the most practical and least leaky). Sections of this AB Blanket are lighter then air and fly in atmosphere. They can be made on a flat area (serving as an assembly area) and delivered by dirigible or helicopter to station at altitude over the city. Here they connect to the already assembled AB Blanket subassemblies, cover the city in an AB Dome and protect it from bad weather, chemical, biological and radioactive fallout or particulates. After finish of dome building the light gas can be changed by air.

Two projects for Manhattan (NY, USA) and Moscow (Russia) are targets for a sample computation.

**Key words:** Dome for city, blanket for city, greenhouse, regional control of weather, protection of cities from chemical, biological and radioactive weapons.
*Presented to http://arxive.org on 3 February 2009.

## Introduction

**Idea**. The inflatable transparent thin film AB Dome offered and developed by author in [1-15] is a good means for converting a city or region into a subtropical garden with excellent weather, obtainable clean water from condensation (and avoided evaporation), saved energy for heating houses (in cold regions), reflecting energy for cooling houses (in hot regions), protection of city from chemical, bacterial, radioactive weapons in war time, even the provision of electricity etc.

However, the author did not describe the method – by which we can cover a city, forest or other obstacle-laden region by thin film.

This article suggests a method for covering the city and any surface which is neither flat nor obstruction free by thin film which insulates the city from outer environment, Earth's atmospheric instabilities, cold winter, strong wind, rain, hot weather and so on.

This new subassembly method of building an inflatable dome is named by the author 'AB-Blanket'. This idea is to design from a transparent double film a blanket, with the internal pockets or space filled by light gas (methane, hydrogen, helium). Subassemblies of the AB Blanket are lighter than air and fly in atmosphere. They can be made in a factory, spread on a flat area, filled by gas to float upwards, and delivered by dirigible or helicopter to a sky over the city. Here they are connected to the AB Dome in building and as additional AB Blankets are brought into place, they cover the city and are sealed together. After finish of dome building the light gas can be changed by air. The film will be supported by small additional air pressure into Dome.

**Information about Earth's megacities.** A megacity is usually defined as a metropolitan area with a total population in excess of 10 million people. Some definitions also set a minimum level for population density (at least 2,000 persons/square km). Megacities can be distinguished from global cities by their rapid growth, new forms of spatial density of population, formal and informal economics. A megacity can be a single metropolitan area or two or more metropolitan areas that converge upon one another. The terms *megapolis* and *megalopolis* are sometimes used synonymously with *megacity*.

In 1800 only 3% of the world's population lived in cities. 47% did by the end of the twentieth century. In 1950, there were 83 cities with populations exceeding one million; but by 2007, this had risen to 468 agglomerations of more than one million. If the trend continues, the world's urban population will double every 38 years, say researchers. The UN forecasts that today's urban population of 3.2 billion will rise to nearly 5 billion by 2030, when three out of five people will live in cities.

The increase will be most dramatic in the poorest and least-urbanised continents, Asia and Africa. Surveys and projections indicate that all urban growth over the next 25 years will be in developing countries. One billion people, one-sixth of the world's population, now live in shanty towns,

By 2030, over 2 billion people in the world will be living in slums. Already over 90% of the urban population of Ethiopia, Malawi and Uganda, three of the world's most rural countries, live in slums.

In 2000, there were 18 megacities – conurbations such as Tokyo, New York City, Los Angeles, Mexico City, Buenos Aires, Mumbai (then Bombay), São Paulo, Karachi that have populations in excess of 10 million inhabitants. Greater Tokyo already has 35 million, which is greater than the entire population of Canada.

By 2025, according to the *Far Eastern Economic Review*, Asia alone will have at least 10 megacities, including Jakarta, Indonesia (24.9 million people), Dhaka, Bangladesh (26 million), Karachi, Pakistan (26.5 million), Shanghai (27 million) and Mumbai (33 million). Lagos, Nigeria has grown from 300,000 in 1950 to an estimated 15 million today, and the Nigerian government estimates that the city will have expanded to 25 million residents by 2015. Chinese experts forecast that Chinese cities will contain 800 million people by 2020.

In 1950, New York was the only urban area with a population of over 10 million. Geographers have identified 25 such areas as of October 2005, as compared with 19 megacities in 2004 and only nine in 1985. This increase has happened as the world's population moves towards the high (75–85%) urbanization levels of North America and Western Europe. The 1990 census marked the first time the majority of US citizens lived in cities with over 1 million inhabitants.

In the 2000s, the largest megacity is the Greater Tokyo Area. The population of this urban agglomeration includes areas such as Yokohama and Kawasaki, and is estimated to be between 35 and 36 million. This variation in estimates can be accounted for by different definitions of what the area encompasses. While the prefectures of Tokyo, Chiba, Kanagawa, and Saitama are commonly included in statistical information, the Japan Statistics Bureau only includes the area within 50 kilometers of the Tokyo Metropolitan Government Offices in Shinjuku, thus arriving at a smaller population estimate. A characteristic issue of megacities is the difficulty in defining their outer limits and accurately estimating the population.

## Description of Innovations

Our design of the dome from levitated AB Blanket sections is presented in Fig.1 that includes the thin inflated film plate parts. The innovations are listed here: (1) the construction is gas-inflatable; (2) each part is fabricated with very thin, transparent film (thickness is 0.05 to 0.2 mm) having controlled clarity (option); (3) the enclosing film has two conductivity layers plus a liquid crystal layer between them which changes its clarity, color and reflectivity under an electric voltage (option); (4) The space between double film is filled a light gas (for example: methane, hydrogen or helium). The air pressure inside the dome is more than the external atmosphere also for protection from outer wind, snow and ice.

The film (textile) may be conventional (and very cheap) or advanced with realtime controlled clarity for cold and hot regions.

The city AB Dome, constructed by means of these AB Blankets, allows getting clean water from rain for drinking, washing and watering which will often be enough for a city population except in case of extreme density (We shall see this for our calculations in the case of Manhattan, below). This water collected at high altitude (Blanket conventionally located at 100 – 500 m) may produce electric energy by hydro-electric generators located at Earth's surface. Wind generators located at high altitude (at Blanket surface) can produce electric energy. Such an AB Dome saves a lot of energy (fuel) for house heating in winter time and cooling in summer time.

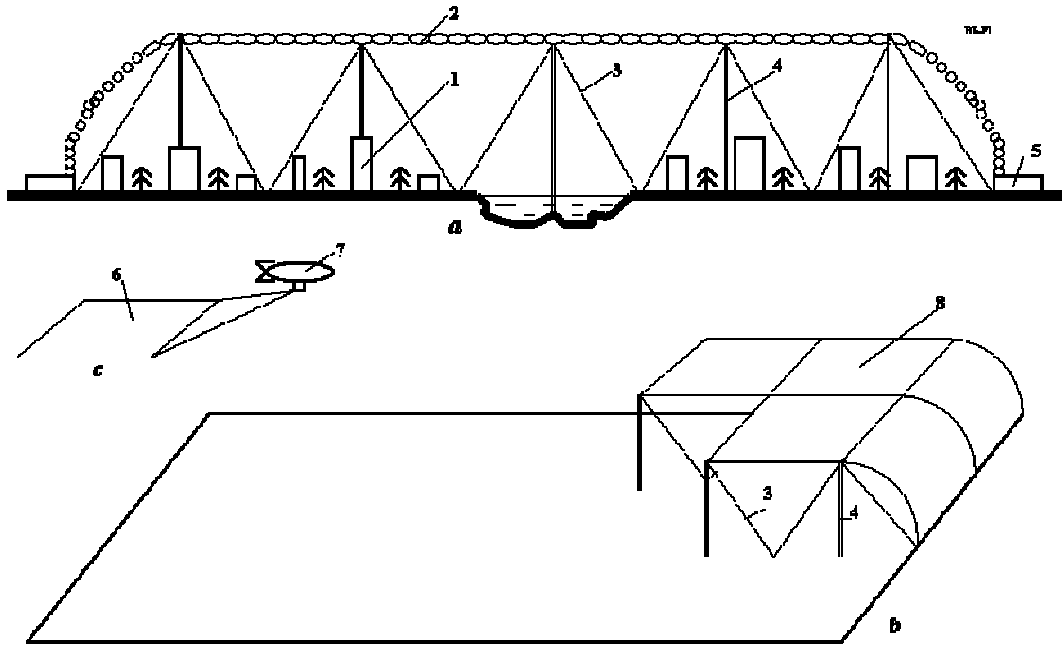

**Fig.1**. (*a*). Design of AB Blanket from the transparent film over city and (*b*) building the AB Dome from parts of Blanket. *Notations:* 1 – city; 2 – AB-Blanket; 3 – bracing wire (support cable); 4 – tubes for rain water, for lifting gas, signalization and control; 5 – enter. Exit and ventilator; 6 – part of Blanket; 7 – dirigible; 8 – building the Blanket.

Detail design of Blanket section is shown in fig.2. Every section contains cylindrical tubes filled a light gas, has margins (explained later in Discussion), has windows which can be open and closed (a full section may be window), connected to Earth's surface by water tube, tube for pumping gas, bracing gables and signal and control wires.

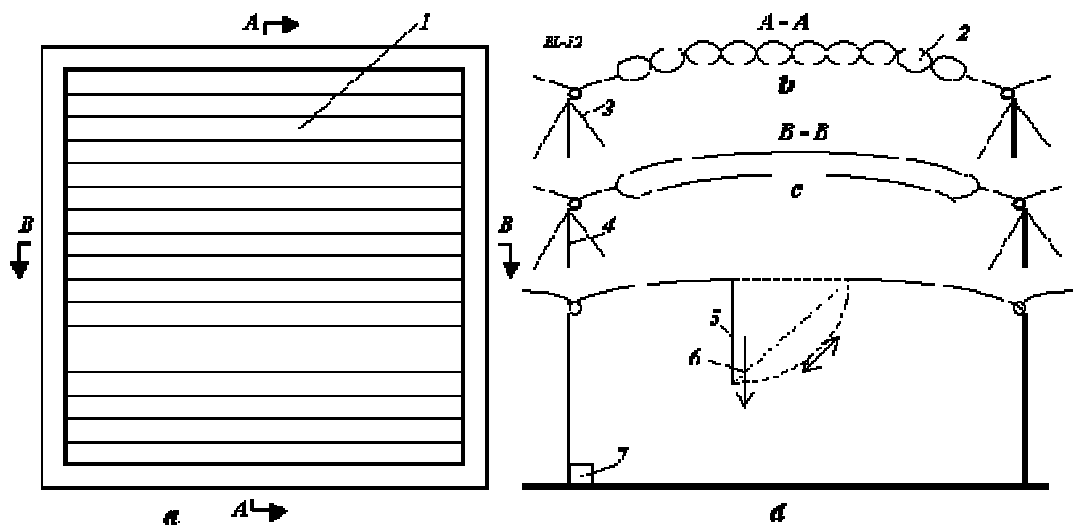

**Fig.2**. Design of AB Blanket section. (***a***) Typical section of Blanket (top view); (***b***) Cross-section A-A of Blanket; (***c***) Cross-section B-B of Blanket; (***d***) Typical section of Blanket (side view). *Notations:* 1 – part of Blanket; 2 – light lift gas (for example: methane, hydrogen or helium); 3 – bracing wire (support cable); 4 – tubes for rain water, for lifting gas, signalization and control; 5 – cover of windows; 6 – snow, ice; 7 – hydro-electric generator, air pump.

Fig. 3 illustrates the advanced thin transparent control Blanket cover we envision. The inflated textile shell—technical "textiles" can be woven or non-woven (films)—embodies the innovations listed: (1) the film is thin, approximately 0.05 to 0.3 mm. A film this thin has never before been used in a major building; (2) the film has two strong nets, with a mesh of about $0.1 \times 0.1$ m and $a = 1 \times 1$ m, the threads are about 0.3 - 1 mm for a small mesh and about 1 - 2 mm for a big mesh. The net prevents the watertight and airtight film covering from being damaged by vibration; (3) the film incorporates a tiny electrically conductive wire net with a mesh about $0.1 \times 0.1$ m and a line width of about 100 μ and a thickness near 10 μ. The wire net is electric (voltage) control conductor. It can inform the dome maintenance engineers concerning the place and size of film damage (tears, rips, etc.); (4) the film has twin-layered with the gap — $c = 1$-$3$ m and $b = 3$-$6$ m—between film layers for heat insulation. In polar (and hot) regions this multi-layered covering is the main means for heat isolation and puncture of one of the layers won't cause a loss of shape because the second film layer is unaffected by holing; (5) the airspace in the dome's covering can be partitioned, either hermetically or not; and (6) part of the covering can have a very thin shiny aluminum coating that is about 1μ (micron) for reflection of unnecessary solar radiation in equatorial or collect additional solar radiation in the polar regions [2].

The town cover may be used as a screen for projection of pictures, films and advertising on the cover at night time. In the case of Manhattan this alone might pay for it!

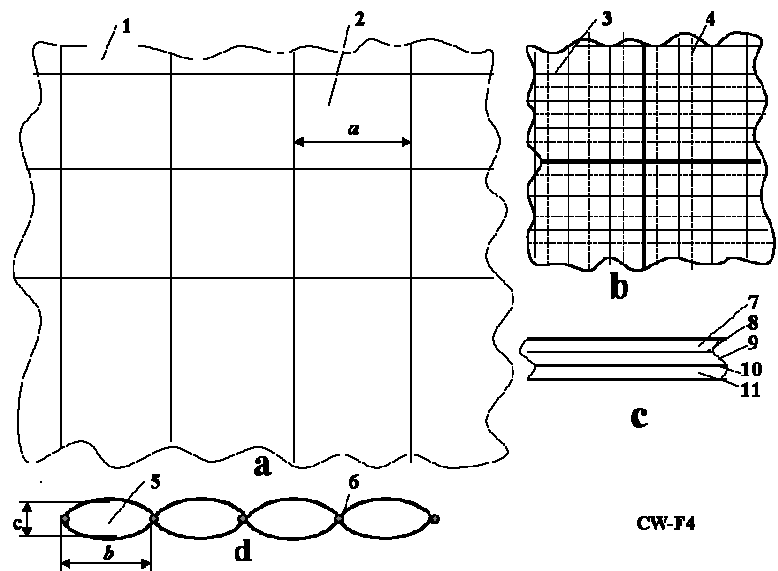

**Fig.3.** Design of advanced covering membrane. *Notations*: (**a**) Big fragment of cover with controlled clarity (reflectivity, carrying capacity) and heat conductivity; (**b**) Small fragment of cover; (**c**) Cross-section of cover (film) having 5 layers; (**d**) Longitudinal cross-section of cover; 1 - cover; 2 -mesh; 3 - small mesh; 4 - thin electric net; 5 - cell of cover; 6 – margins and wires; 7 - transparent dielectric layer; 8 - conducting layer (about 1 - 3 μ); 9 - liquid crystal layer (about 10 - 100 μ); 10 - conducting layer; and 11 - transparent dielectric layer. Common thickness is 0.1 - 0.5 mm. Control voltage is 5 - 10 V.

**Brif information about advanced cover film.** Our advanced Blanket cover (film) has 5 layers (fig. 3c): transparent dielectric layer, conducting layer (about 1 - 3 μ), liquid crystal layer (about 10 - 100 μ), conducting layer (for example, $SnO_2$), and transparent dielectric layer. Common thickness is 0.3 - 1 mm. Control voltage is 5 - 10 V. This film may be produced by industry relatively cheaply.

**1. Liquid crystals** (LC) are substances that exhibit a phase of matter that has properties between those of a conventional liquid, and those of a solid crystal.

Liquid crystals find wide use in liquid crystal displays (LCD), which rely on the optical properties of certain liquid crystalline molecules in the presence or absence of an electric field. The electric field can be used to make a pixel switch between clear or dark on command. Color LCD systems use the same technique, with color filters used to generate red, green, and blue pixels. Similar principles can be used to make other liquid crystal based optical devices. Liquid crystal in fluid form is used to detect electrically generated hot spots for failure analysis in the semiconductor industry. Liquid crystal memory units with extensive capacity were used in Space Shuttle navigation equipment. It is also worth noting that many common fluids are in fact liquid crystals. Soap, for instance, is a liquid crystal, and forms a variety of LC phases depending on its concentration in water.

The conventional controlled clarity (transparency) film reflects superfluous energy back to space if too much. If film has solar cells it may converts part of the superfluous solar energy into electricity.

**2. Transparency**. In optics, transparency is the material property of allowing light to pass through. Though transparency usually refers to visible light in common usage, it may correctly be used to refer to any type of radiation. Examples of transparent materials are air and some other gases, liquids such as water, most glasses, and plastics such as Perspex and Pyrex. Where the degree of transparency varies according to the wavelength of the light. From electrodynamics it results that only a vacuum is really transparent in the strict meaning, any matter has a certain absorption for electromagnetic waves. There are transparent glass walls that can be made opaque by the application of an electric charge, a technology known as electrochromics.Certain crystals are transparent because there are straight lines through the crystal structure. Light passes unobstructed along these lines. There is a complicated theory "predicting" (calculating) absorption and its spectral dependence of different materials. The optic glass has transparance about 95% of light (visible) radiation. The transparancy dipents fron thickness and may be very high for thin film.

**3. Electrochromism** is the phenomenon displayed by some chemical species of reversibly changing color when a burst of charge is applied.

One good example of an electrochromic material is polyaniline which can be formed either by the electrochemical or chemical oxidation of aniline. If an electrode is immersed in hydrochloric acid which contains a small concentration of aniline, then a film of polyaniline can be grown on the electrode. Depending on the redox state, polyaniline can either be pale yellow or dark green/black. Other electrochromic materials that have found technological application include the viologens and polyoxotungstates. Other electrochromic materials include tungsten oxide ($WO_3$), which is the main chemical used in the production of electrochromic windows or smart windows.

As the color change is persistent and energy need only be applied to effect a change, electrochromic materials are used to control the amount of light and heat allowed to pass through windows ("smart windows"), and has also been applied in the automobile industry to automatically tint rear-view mirrors in various lighting conditions. Viologen is used in conjunction with titanium dioxide ($TiO_2$) in the creation of small digital displays. It is hoped that these will replace LCDs as the viologen (which is typically dark blue) has a high contrast to the bright color of the titanium white, therefore providing a high visibility of the display.

### 3. THEORY AND COMPUTATIONS OF THE AB BLANKET

1. **Lift force of Blanket.** The specific lift force of Blanket is computed by equation:

$$L = g(q_a - q_g)V , \qquad (1)$$

where $L$ is lift force, N; $g$ = 9.81 m/s$^2$ is gravity; $q_a$= 1.225 kg/m$^3$ is an air density for standard condition ($T$ = 15$^o$C); $q_g < q_a$ is density of lift light gas. For methane $q_g$ = 0.72 kg/m$^3$, hydrogen $q_g$ = 0.09 kg/m$^3$, helium $q_g$ = 0.18 kg/m$^3$; $V$ is volume of Blanket, m$^3$. For example, the section 100×100m of the Blanket filled by methane (the cheapest light gas) having the average thickness 3 m has the lift force 15 N/m$^2$ or 150,000N = 15 tons.

**2. The weight (mass) of film** may be computed by equation

$$W = \gamma \delta S, \qquad (2)$$

where $W$ is weight of film, kg; $\gamma$ is specific density of film (usually about $\gamma = 1500 \div 1800$ kg/m$^3$); $\delta$ is thickness, m; $S$ is area, m$^2$. For example, the double film of thickness $\delta = 0.05$ mm has weight $W = 0.15$ kg/m$^2$. The section 100×100m of the Blanket has weight 1500 kg = 1.5 tons.

**3. Weight (mass) of support cable** (bracing wire) is computed by equation:

$$W_c = \gamma_c \frac{hLS}{\sigma}, \qquad (3)$$

where $W_c$ is weight of support cable, kg; $\gamma_c$ is specific density of film (usually about $\gamma_c = 1800$ kg/m$^3$); $\sigma$ is safety density of cable, N/m$^2$. For cable from artificial fiber $\sigma = 100 \div 150$ kg/mm$^2$ = $(1 \div 1.5) \times 10^9$ N/m$^2$. For example, for $\sigma = 100$ kg/mm$^2$, $h = 500$ m, $L = 10$ N/m$^2$, $W_c = 0.009$ kg/m$^2$. However, if additional air pressure into dome is high, for example, lift force $L = 1000$ N/m$^2$ (air pressure $P = 0.01$ atm – 0.01 bar), the cable weight may reach 0.9 kg/m$^2$. That may be requested in a storm weather when outer wind and wind dynamic pressure is high.

As wind flows over and around a fully exposed, nearly completely sealed inflated dome, the weather affecting the external film on the windward side must endure positive air pressures as the wind stagnates. Simultaneously, low air pressure eddies will be present on the leeward side of the dome. In other words, air pressure gradients caused by air density differences on different parts of the sheltering dome's envelope is characterized as the "buoyancy effect". The buoyancy effect will be greatest during the coldest weather when the dome is heated and the temperature difference between its interior and exterior are greatest. In extremely cold climates, such as the Arctic and Antarctica, the buoyancy effect tends to dominate dome pressurization, causing the Blanket to require reliable anchoring.

**4. The wind dynamic pressure** is computed by equation

$$p_d = \frac{\rho V^2}{2}, \qquad (4)$$

where $p_d$ is wind dynamic pressure, N/m$^2$; $\rho$ is air density, for altitude $H = 0$ the $\rho = 1.225$ kg/m$^3$; $V$ is wind speed, m/s. The computation is presented in fig.4.

The small overpressure of 0.01 atm forced into the AB-Dome to inflate it produces force $p = 1000$ N/m$^2$. That is greater than the dynamic pressure (740 N/m$^2$) of very strong wind $V = 35$ m/s (126 km/hour). If it is necessary we can increase the internal pressure by some times if needed for very exceptional storms.

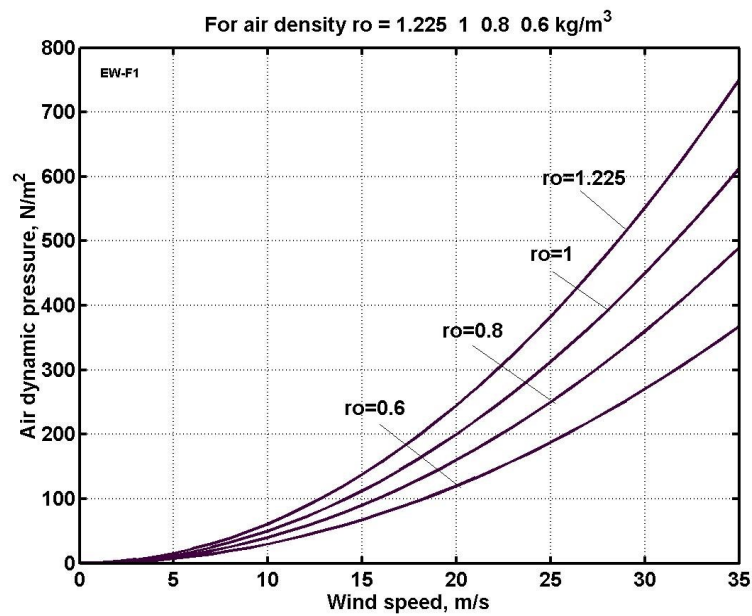

**Fig. 4.** Wind dynamic pressure versus wind speed and air density $\rho$. The $ro = 0.6$ is for $H \approx 6$ km.

**5. The thickness of the dome envelope**, its sheltering shell of film, is computed by formulas (from equation for tensile strength):

$$\delta_1 = \frac{Rp}{2\sigma}, \quad \delta_2 = \frac{Rp}{\sigma}, \tag{5}$$

where $\delta_1$ is the film thickness for a spherical dome, m; $\delta_2$ is the film thickness for a cylindrical dome, m; $R$ is radius of dome, m; $p$ is additional pressure into the dome, N/m$^2$; $\sigma$ is safety tensile stress of film, N/m$^2$.

For example, compute the film thickness for dome having radius $R = 50$ m, additional internal air pressure $p = 0.01$ atm ($p = 1000$ N/m$^2$), safety tensile stress $\sigma = 50$ kg/mm$^2$ ($\sigma = 5 \times 10^8$ N/m$^2$), cylindrical dome.

$$\delta = \frac{50 \times 1000}{5 \times 10^8} = 0.0001 m = 0.1 \; mm \tag{5'}$$

**6. Solar radiation.** Our basic computed equations, below, are derived from a Russian-language textbook [19]. Solar radiation impinging the orbiting Earth is approximately 1400 W/m$^2$. The average Earth reflection by clouds and the sub-aerial surfaces (water, ice and land) is about 0.3. The Earth-atmosphere adsorbs about 0.2 of the Sun's radiation. That means about $q_0 = 700$ W/m$^2$s of solar energy (heat) reaches our planet's surface at the Equator. The solar spectrum is graphed in Fig. 5.

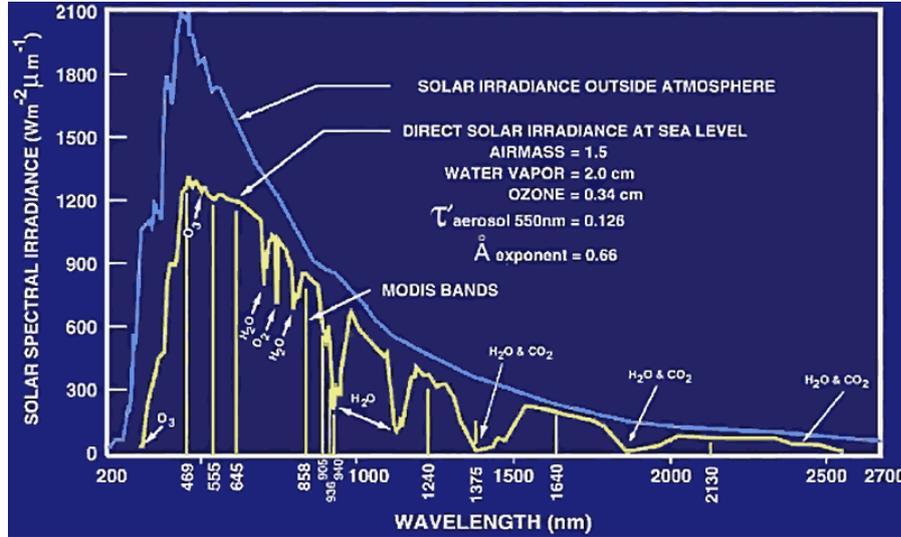

**Fig.5.** Spectrum of solar irradiance outside atmosphere and at sea level with absorption of electromagnetic waves by atmospheric gases. Visible light is 0.4 - 0.8 $\mu$ (400 – 800 nm).

The visible part of the Sun's spectrum is only $\lambda = 0.4$ to $0.8 \; \mu$. Any warm body emits radiation. The emission wavelength depends on the body's temperature. The wavelength of the maximum intensity (see Fig. 5) is governed by the black-body law originated by Max Planck (1858-1947):

$$\lambda_m = \frac{2.9}{T}, \quad [mm], \tag{6}$$

where $T$ is body temperature, °K. For example, if a body has an ideal temperature 20 °C ($T = 293$ °K), the wavelength is $\lambda_m = 9.9 \; \mu$.

The energy emitted by a body may be computed by employment of the Josef Stefan-Ludwig Boltzmann law.

$$E = \varepsilon \sigma_s T^4, \quad [W/m^2], \tag{7}$$

where $\varepsilon$ is coefficient of body blackness ($\varepsilon = 0.03 \div 0.99$ for real bodies), $\sigma_s = 5.67 \times 10^{-8}$ [W/m$^2$·K] Stefan-Boltzmann constant. For example, the absolute black-body ($\varepsilon = 1$) emits (at $T = 293$ °C) the energy $E = 418$ W/m$^2$.

Amount of the maximum solar heat flow at 1 m$^2$ per 1 second of Earth surface is

$$q = q_o \cos(\varphi \pm \theta) \quad [W/m^2], \tag{8}$$

where $\varphi$ is Earth longevity, $\theta$ is angle between projection of Earth polar axis to the plate which is perpendicular to the ecliptic plate and contains the line Sun-Earth and the perpendicular to ecliptic plate. The sign "+" signifies Summer and the "-" signifies Winter, $q_o \approx 700$ W/m² is the annual average solar heat flow to Earth at equator corrected for Earth reflectance.

This angle is changed during a year and may be estimated for the Arctic by the following the first approximation equation:

$$\theta = \theta_m \cos\omega, \quad \text{where} \quad \omega = 2\pi \frac{N}{364}, \qquad (9)$$

where $\theta_m$ is maximum $\theta$, $|\theta_m| = 23.5° = 0.41$ radian; $N$ is number of day in a year. The computations for Summer and Winter are presented in fig.6.

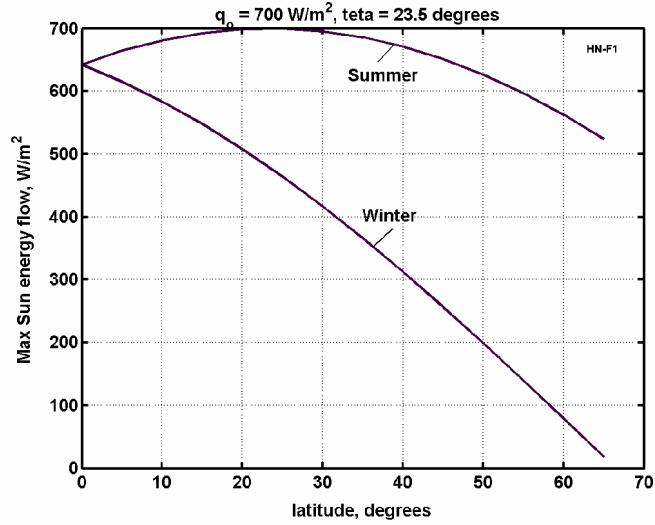

**Fig.6.** Maximum Sun radiation flow at Earth surface as function of Earth latitude and season.

The heat flow for a hemisphere having reflector (fig.1) at noon may be computed by equation

$$q = c_1 q_0 [\cos(\varphi - \theta) + S \sin(\varphi + \theta)], \qquad (10)$$

where $S$ is fraction (relative) area of reflector to service area of "Evergreen" dome. Usually $S = 0.5$; $c_1$ is film transparency coefficient ($c_1 \approx 0.9 - 0.95$).

The daily average solar irradiation (energy) is calculated by equation

$$Q = 86400 c q t, \quad \text{where} \quad t = 0.5(1 + \tan\varphi \tan\theta), \quad |\tan\varphi \tan\theta| \leq 1, \qquad (11)$$

where $c$ is daily average heat flow coefficient, $c \approx 0.5$; $t$ is relative daylight time, $86400 = 24 \times 60 \times 60$ is number of seconds in a day.

The computation for relative daily light period is presented in Fig. 7.

The heat loss flow per 1 m² of dome film cover by convection and heat conduction is (see [19]):

$$q = k(t_1 - t_2), \quad \text{where} \quad k = \frac{1}{1/\alpha_1 + \sum_i \delta_i / \lambda_i + 1/\alpha_2}, \qquad (12)$$

where $k$ is heat transfer coefficient, W/m²·K; $t_{1,2}$ are temperatures of the inter and outer multi-layers of the heat insulators, °C; $\alpha_{1,2}$ are convention coefficients of the inter and outer multi-layers of heat insulators ($\alpha = 30 \div 100$), W/m²K; $\delta_i$ are thickness of insulator layers; $\lambda_i$ are coefficients of heat transfer of insulator layers (see Table 1), m; $t_{1,2}$ are temperatures of initial and final layers °C.

The radiation heat flow per 1 m²s of the service area computed by equations (7):

$$q = C_r \left[ \left(\frac{T_1}{100}\right)^4 - \left(\frac{T_2}{100}\right)^4 \right], \quad \text{where} \quad C_r = \frac{c_s}{1/\varepsilon_1 + 1/\varepsilon_2 - 1}, \quad c_s = 5.67 \text{ [W/m}^2\text{K}^4\text{]}, \qquad (13)$$

where $C_r$ is general radiation coefficient, $\varepsilon$ are black body rate (Emittance) of plates (see Table 2); $T$ is temperatures of plates, °K.

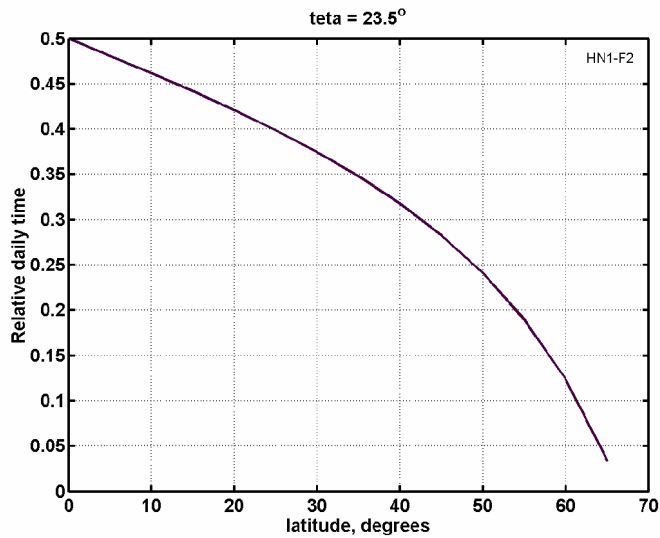

**Fig.7**. Relative daily light time relative to Earth latitude.

The radiation flow across a set of the heat reflector plates is computed by equation

$$q = 0.5 \frac{C'_r}{C_r} q_r, \qquad (14)$$

where $C'_r$ is computed by equation (8) between plate and reflector.
The data of some construction materials is found in Table 1, 2.

**Table 1**. [19], p.331. Heat Transfer.

| Material | Density, kg/m$^3$ | Thermal conductivity, $\lambda$, W/m·°C | Heat capacity, kJ/kg·°C |
|---|---|---|---|
| Concrete | 2300 | 1.279 | 1.13 |
| Baked brick | 1800 | 0.758 | 0.879 |
| Ice | 920 | 2.25 | 2.26 |
| Snow | 560 | 0.465 | 2.09 |
| Glass | 2500 | 0.744 | 0.67 |
| Steel | 7900 | 45 | 0.461 |
| Air | 1.225 | 0.0244 | 1 |

As the reader will see, the air layer is the best heat insulator. We do not limit its thickness $\delta$.

**Table 2**. Nacshekin (1969), p. 465. Emittance, $\varepsilon$ (Emissivity)

| Material | Temperature, $T$ °C | Emittance, $\varepsilon$ |
|---|---|---|
| Bright Aluminum | 50 ÷ 500 °C | 0.04 - 0.06 |
| Bright copper | 20 ÷ 350 °C | 0.02 |
| Steel | 50 °C | 0.56 |
| Asbestos board | 20 °C | 0.96 |
| Glass | 20 ÷ 100 °C | 0.91 - 0.94 |
| Baked brick | 20 °C | 0.88 - 0.93 |
| Tree | 20 °C | 0.8 - 0.9 |
| Black vanish | 40 ÷ 100 °C | 0.96 – 0.98 |
| Tin | 20 °C | 0.28 |

As the reader will notice, the shiny aluminum louver coating is an excellent mean jalousie (louvered window, providing a similar service to a Venetian blind) which serves against radiation losses from the dome.

The general radiation heat $Q$ computes by equation [11]. Equations [6] – [14] allow computation of the heat balance and comparison of incoming heat (gain) and outgoing heat (loss).
The computations of heat balance of a dome of any size in the coldest wintertime of the Polar Regions are presented in Fig. 8.

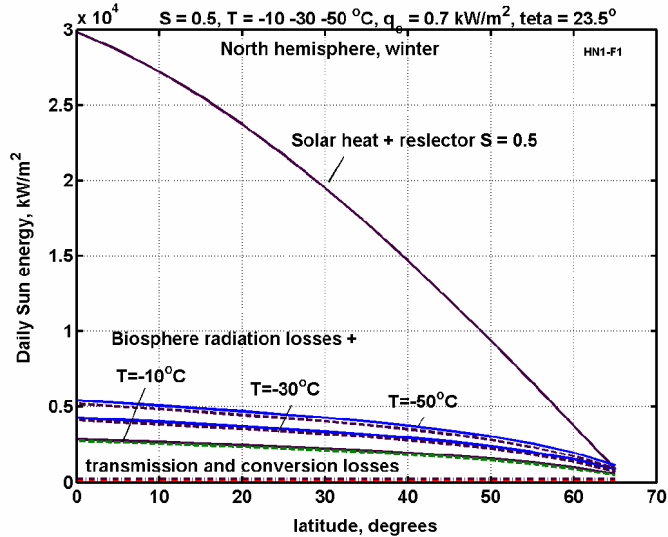

**Fig. 8.** Daily heat balance through 1 m² of dome during coldest winter day versus Earth's latitude (North hemisphere example). Data used for computations (see Eq. (6) - (14)): temperature inside of dome is $t_1$= +20 °C, outside are $t_2$ = -10, -30, -50 °C; reflectivity coefficient of mirror is $c_2$= 0.9; coefficient transparency of film is $c_1$ = 0.9; convectively coefficients are $\alpha_1$= $\alpha_2$ = 30; thickness of film layers are $\delta_1$= $\delta_2$ =0.0001 m; thickness of air layer is $\delta$ = 1 m; coefficient of film heat transfer is $\lambda_1$= $\lambda_3$ = 0.75, for air $\lambda_2$ = 0.0244; ratio of cover blackness $\varepsilon_1$= $\varepsilon_3$ = 0.9, for louvers $\varepsilon_2$ = 0.05.

The heat from combusted fuel is found by equation
$$Q = c_t m/\eta, \qquad (15)$$
where $c_t$ is heat rate of fuel [J/kg]; $c_t$ = 40 MJ/kg for liquid oil fuel; $m$ is fuel mass, kg; $\eta$ is efficiency of heater, $\eta$ = 0.5 - 0.8.

In Fig. 8 the alert reader has noticed: the daily heat loss is about the solar heat in the very coldest Winter day when a dome located above 60⁰ North or South Latitude and the outside air temperature is –50 °C.

**7. Properties and Cost of material.** The cost some material are presented in Table 3 (2005-2007). Properties are in Table 4. Some difference in the tensile stress and density are result the difference sources, models and trademarks.

**Table 3**. Average cost of material (2005-2007)

| Material | Tensile stress, MPa | Density, g/cm³ | Cost USD $/kg |
|---|---|---|---|
| **Fibers:** | | | |
| Glass | 3500 | 2.45 | 0.7 |
| Kevlar 49, 29 | 2800 | 1.47 | 4.5 |
| PBO Zylon AS | 5800 | 1.54 | 15 |
| PBO Zylon HM | 5800 | 1.56 | 15 |
| Boron | 3500 | 2.45 | 54 |
| SIC | 3395 | 3.2 | 75 |
| Saffil (5% SiO₂+Al₂O₃) | 1500 | 3.3 | 2.5 |
| **Matrices:** | | | |
| Polyester | 35 | 1,38 | 2 |
| Polyvinyl | 65 | 1.5 | 3 |
| Aluminum | 74-550 | 2.71 | 2 |
| Titanum | 238-1500 | 4.51 | 18 |
| Borosilicate glass | 90 | 2.23 | 0.5 |
| Plastic | 40-200 | 1.5-3 | 2 - 6 |
| **Materials:** | | | |
| Steel | 500 - 2500 | 7.9 | 0.7 - 1 |
| Concrete | - | 2.5 | 0.05 |
| Cement (2000) | - | 2.5 | 0.06-0.07 |
| Melted Basalt | 35 | 2.93 | 0.005 |

Table 4. Material properties

| Material | Tensile strength kg/mm$^2$ | Density g/cm$^3$ | | Tensile strength kg/mm$^2$ | Density g/cm$^3$ |
|---|---|---|---|---|---|
| **Whiskers** | | | **Fibers** | | |
| AlB$_{12}$ | 2650 | 2.6 | QC-8805 | 620 | 1.95 |
| B | 2500 | 2.3 | TM9 | 600 | 1.79 |
| B$_4$C | 2800 | 2.5 | Allien 1 | 580 | 1.56 |
| TiB$_2$ | 3370 | 4.5 | Allien 2 | 300 | 0.97 |
| SiC | 1380-4140 | 3.22 | Kevlar or Twaron | 362 | 1.44 |
| **Material** | | | Dynecta or Spectra | 230-350 | 0.97 |
| Steel prestressing strands | 186 | 7.8 | Vectran | 283-334 | 0.97 |
| Steel Piano wire | 220-248 | | E-Glass | 347 | 2.57 |
| Steel A514 | 76 | 7.8 | S-Glass | 471 | 2.48 |
| Aluminum alloy | 45.5 | 2.7 | Basalt fiber | 484 | 2.7 |
| Titanium alloy | 90 | 4.51 | Carbon fiber | 565 | 1,75 |
| Polypropylene | 2-8 | 0.91 | Carbon nanotubes | 6200 | 1.34 |

Source: Howatsom A.N., Engineering Tables and Data, p.41.

**8. Closed-loop water cycle.** The closed Dome allows creating a closed loop cycle, when vapor water in the day time will returns as condensation or dripping rain in the night time. A reader can derive the equations below from well-known physical laws Nacshekin [19](1969). Therefore, the author does not give detailed explanations of these.

**Amount of water in atmosphere**. Amount of water in atmosphere depends upon temperature and humidity. For relative humidity 100%, the maximum partial pressure of water vapor for pressure 1 atm is shown in Table 5.

Table 5. Maximum partial pressure of water vapor in atmosphere for given air temperature (pressure is 1 atm)

| $t$, C | -10 | 0 | 10 | 20 | 30 | 40 | 50 | 60 | 70 | 80 | 90 | 100 |
|---|---|---|---|---|---|---|---|---|---|---|---|---|
| $p$, kPa | 0.287 | 0.611 | 1.22 | 2.33 | 4.27 | 7.33 | 12.3 | 19.9 | 30.9 | 49.7 | 70.1 | 101 |

The amount of water in 1 m$^3$ of air may be computed by equation
$$m_W = 0.00625\,[p(t_2)h - p(t_1)], \tag{16}$$
where $m_W$ is mass of water, kg in 1 m$^3$ of air; $p(t)$ is vapor (steam) pressure from Table 4, relative $h = 0 \div 1$ is relative humidity. The computation of equation (16) is presented in fig.9. Typical relative humidity of atmosphere air is 0.5 - 1.

**Computation of closed-loop water cycle.** Assume the maximum safe temperature is achieved in the daytime. When dome reaches the maximum (or given) temperature, the control system fills with air the space 5 (Fig.3) between double–layers of the film cover. That protects the inside part of the dome from further heating by outer (atmospheric) hot air. The control system decreases also the solar radiation input, increasing reflectivity of the liquid crystal layer of the film cover. That way, we can support a constant temperature inside the dome.

The **heating** of the dome in the daytime may be computed by equations:

$$q(t) = q_0 \sin(\pi t / t_d), \quad dQ = q(t)dt, \quad Q = \int_0^{t_d} dQ, \quad Q(0) = 0, \quad M_w = \int_0^{t_d} a\, dT,$$

$$dT = \frac{dQ}{C_{p1}\rho_1\delta_1 + C_{p2}\rho_2 H + rHa}, \quad a = 10^{-5}(5.28T + 2), \quad T = \int_0^{t_d} dT, \quad T(0) = T_{min}, \tag{17}$$

where $q$ is heat flow, J/m²s; $q_o$ is maximal Sun heat flow in daily time, $q_o \approx 100 \div 900$, J/m²s; $t$ is time, s; $t_d$ is daily (Sun) time, s; $Q$ is heat, J; $T$ is temperature in dome (air, soil), °C; $C_{p1}$ is heat capacity of soil, $C_{p1} \approx 1000$ J/kg; $C_{p2} \approx 1000$ J/kg is heat capacity of air; $\delta_1 \approx 0.1$ m is thickness of heating soil; $\rho_1 \approx 1000$ kg/m³ is density of the soil; $\rho_2 \approx 1.225$ kg/m³ is density of the air; $H$ is thickness of air (height of cover), $H \approx 5 \div 300$ m; $r = 2,260,000$ J/kg is evaporation heat, $a$ is coefficient of evaporation; $M_w$ is mass of evaporation water, kg/m³; $T_{min}$ is minimal temperature into dome after night, °C.

The convective (conductive) cooling of dome at night time may be computed as below

$$q_t = k(T_{min} - T(t)), \quad \text{where} \quad k = \frac{1}{1/\alpha_1 + \sum_i \delta_i / \lambda_i + 1/\alpha_2} \tag{18}$$

where $q_t$ is heat flow through the dome cover by convective heat transfer, J/m²s or W/m²; see the other notation in Eq. (12). We take $\delta = 0$ in night time (through active control of the film).

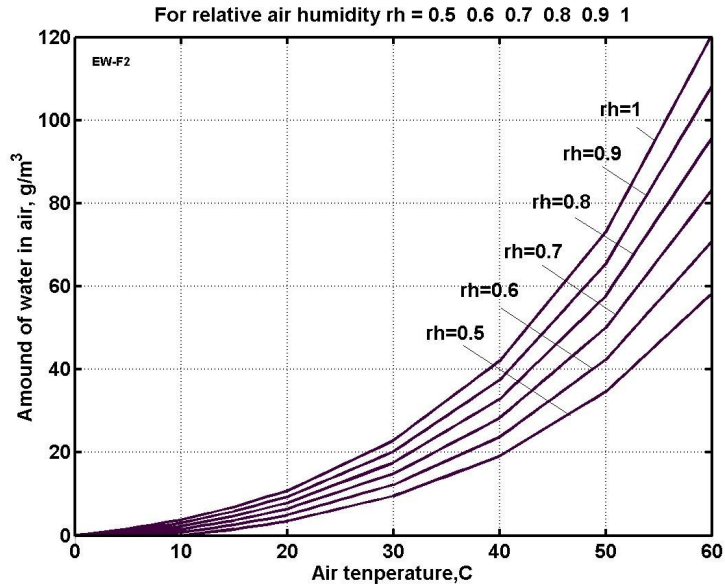

**Fig. 9**. Amount of water in 1 m³ of air versus air temperature and relative humidity (rh). $t_1 = 0$ °C.

The radiation heat flow $q_r$ (from dome to night sky, radiation cooling) may be estimated by equations (10).

$$q_r = C_r\left[\left(\frac{T_{min}}{100}\right)^4 - \left(\frac{T(t)}{100}\right)^4\right], \quad \text{where} \quad C_r = \frac{c_s}{1/\varepsilon_1 + 1/\varepsilon_2 - 1}, \quad c_s = 5.67 \text{ [W/m}^2\text{K}^4\text{]}, \tag{19}$$

where $q_r$ is heat flow through dome cover by radiation heat transfer, J/m²s or W/m²; see the other notation in Eq. (10). We take $\varepsilon = 1$ in night time (through active control of the film).

The other equations are same (17)

$$dQ = [q_t(t) + q_r(t)]dt, \quad Q = \int_0^{t_d} dQ, \quad Q(0) = 0, \quad M_w = \int_0^{t_d} a\, dT,$$

$$dT = \frac{dQ}{C_{p1}\rho_1\delta_1 + C_{p2}\rho_2 H + rHa}, \quad a = 10^{-5}(5{,}28T + 2), \quad T = \int_0^{t_d} dT, \quad T(0) = T_{min}, \tag{20}$$

Let us take the following parameters: $H$ = 135 m, $\alpha$ =70, $\delta$ = 1 m between cover layers, $\lambda$ = 0.0244 for air. Result of computation for given parameter are presented in figs. 10 – 11.

For dome cover height $H$ = 135 m the night precipitation (maximum) is 0.027×135 = 3.67 kg (liter) or 3.67 mm/day (Fig.12). The AB Dome's internal annual precipitation under these conditions is is 1336.6 mm (maximum). If it is not enough, we can increase the height of dome cover. The globally-averaged annual precipitation is about 1000 mm on Earth.

As you see, we can support the same needed temperature in a wide range of latitudes at summer and winter time. That means the covered regions are not hostage to their location upon the Earth's surface (up to latitude $20^o$ -$30^o$), nor Earth's seasons, nor it is dependant upon outside weather. Our design of Dome is not optimal, but rather selected for realistic parameters.

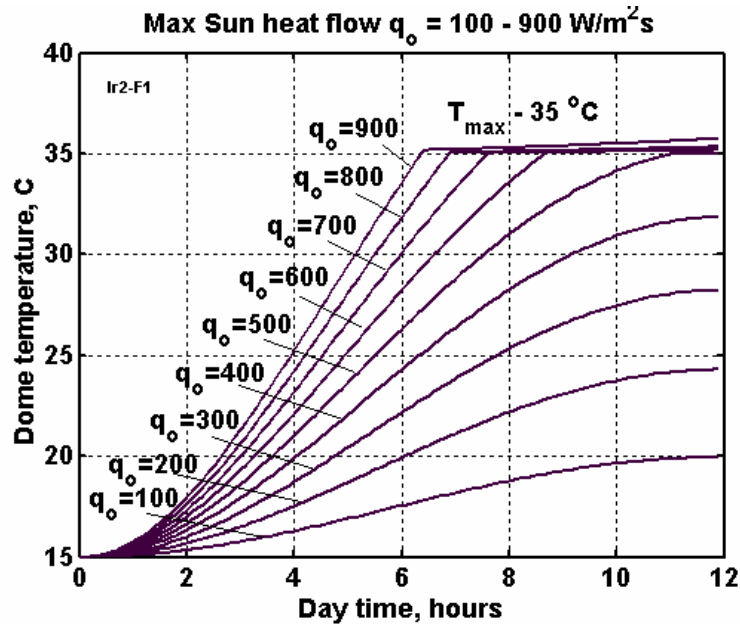

**Fig. 10.** Heating of the dome by solar radiation from the night temperature of $15\,^o$ C to $35\,^o$ C via daily maximal solar radiation (W/m$^2$) for varying daily time. Height of dome film cover equals $H$ = 135 m. The control temperature system limits the maximum internal dome temperature to $35\,^o$ C.

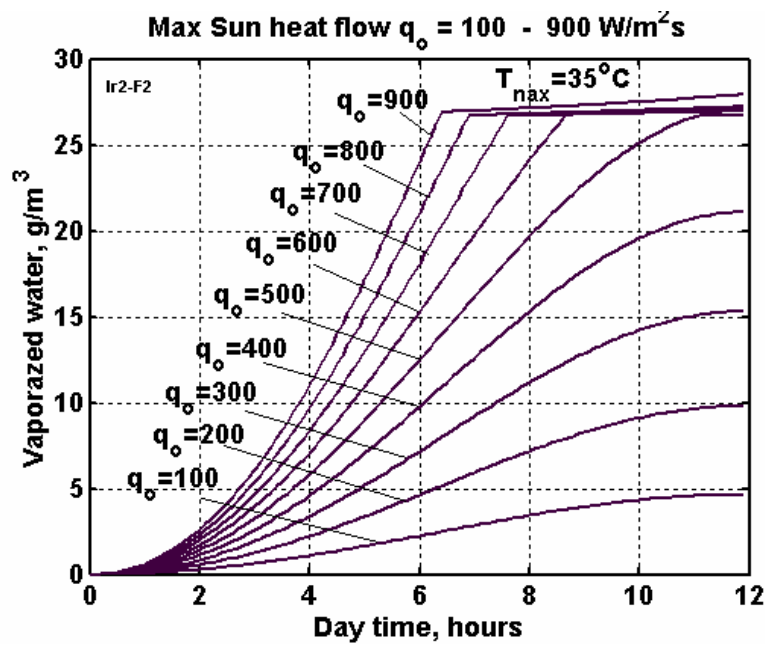

**Fig.11.** Water vaporization for 100% humidity of the air for different maximal solar radiation (W/m²) levels delivered over varying daily time. Height of dome film cover equals $H = 135$ m. The temperature control system limits the maximum internal dome temperature to 35 C.

# Projects

**Project 1. Manhattan** (district of New-York, USA).

**Manhattan Island**, in New York Harbor, is the largest part of the Borough of Manhattan, one of the Five Boroughs which form the City of New York. With a 2007 population of 1,620,867 living in a land area of 22.96 square miles (59.47 km²), New York County is the most densely populated county in the United States at 70,595 residents per square mile (27,267/km²). It is also one of the wealthiest counties in the United States, with a 2005 personal per capita income above $100,000.

Area (land) of Manhattan is 22.96 sq mi (59.5 km², $A \approx 60$ km²), population 1,620,867 inhabitants, density 70,595/sq mi (27,256.9/km²).

Average annual high temperature is 17C (62F), average annual low temperature is 8C (47F). The average high monthly temperature is 30C (July), the average low monthly temperature is -4C (January). Annual rainfall is 1,124 mm.

**Computation and estimation of cost:**

**Film**. Requested area of double film is $A_f = 3 \times 60$ km² = 180 km². If thickness of film is $\delta = 0.1$ mm, specific density $\gamma = 1800$ kg/m³, the mass of film is $M = \gamma \delta A_f = 32,500$ tons or $m = 0.54$ kg/m². If cost of film is $c$ - $2/kg, the total cost of film is $C_f = cM = \$65$ millions or $c_a = \$1.08$/m².

If average thickness of a gas layer inside the AB-Blanket is $\delta = 3$ m, the total volume of gas is $V = \delta A = 1.8 \times 10^8$ m³. One m³ of methane ($CH_4$) has lift force $l = 0.525$ kg/m³ or Blanket of thickness $\delta = 3$ m has lift force $l = 1.575$ kg/m² or the total Blanket lift force is $L = 94.5 \times 10^3$ tons. Cost of methane is $c = \$0.4$/m³, volume is $V = \delta A = 1.8 \times 10^8$ m³. But we did not take in account because after finishing building the AB Dome the methane will be changed for overpressured air. (Thus $72 million in methane would not be kept in inventory, but if the AB-Blankets were each 1% of the final area, neglecting leaks only $720,000 worth of methane would be in play at any one time. With some designs step by step methane replacement with air will be possible (if overpressure support is introduced another way, etc.)

**Support cables**. Let us take an additional air pressure as $p = 0.01$ atm = 1000 N/m², safety tensile stress of artificial fiber $\sigma = 100$ kG/mm², specific density $\gamma = 1800$ kg/m³, $s = 1$ m², and altitude of the Blanket $h = 500$ m. Then needed cross-section of cable is 1 mm² per 1 m² of Blanket and mass of the support cable is $m = \gamma p h/\sigma = 0.9$ kg per 1 m² of Blanket. If cost of fiber is $1/kg, the cost of support cable is $c_c = \$0.9$/m². Total mass of the support cables is 54,000 tons.

The average cost of air and water **tubes and control system** we take $c_t = \$0.5$/m².

The **total cost** of 1 m² material is $C = c_a + c_c + c_t = 1.08 + 0.9 + 0.5 = \$2.48$/m² ≈ $2.5/m² or $150 millions of the USA dollars for Manhattan area. The work will cost about $100 million. *The total cost of Blanket construction for Manhattan is about **$250 million** US dollars.*

The clean (rain) water is received from 1 m² of covered area is 1.1 kL/year. That is enough for the Manhattan population. The possible energy (if we install at extra expense hydro-electric generators and utilize pressure (50 atm) of the rain water) is about 4000 kJ/m² in year. That covers about 15% of city consumption.

Manhattan receives a permanent warm climate and saves a lot of fuel for home heating (decreased pollution of atmosphere) in winter time and save a lot of electric energy for home cooling in the summer time.

**Project 2. Moscow** (Russia)

Area (land) of Moscow is 1,081 km² (417.4 sq mi), population (as of the 2002 Census) 10,470,318 inhabitants, density 9,685.8/km² (25,086.1/sq mi).

Average annual high temperature is 9.1C, average annual low temperature is 2.6C. The average high monthly temperature is 24C (July)(Record is 36.5C), the average low monthly temperature is -8C (January)(Record low is -42.2C). Annual rainfall is 705 mm.

**Estimation.** The full Moscow area is significantly more the Manhattan area (by 18 times) and has less population density (by 3 times). We can cover only the most important central part of Moscow, the place where are located the Government and business offices, tourist hotels, theaters and museums.

If this area equals 60 km$^2$ the cost of construction will be cheaper than $250 million US because the labor cost less (by 3 -5 times) then the USA. But profit from Moscow Blanket may be more then from the Manhattan cover because the weather is colder in Moscow than in New York.

## DISCUSSION

As with any innovative macro-project proposal, the reader will naturally have many questions. We offer brief answers to the most obvious questions our readers are likely to ponder.

(1) *The methane gas is fuel. How about fire protection?*

The AB Blanket is temporarily filled by methane gas for air delivery and period of Dome construction. After finish of dome construction the methane will be changed by air and the Blanket will be supported at altitude by small additional air pressure into AB-Dome.

The second reason: the Blanket contains methane in small separated cylindrical sections (in piece 100×100 m has about 30 these sections, see fig.2) and every piece has special anti-fire margins (fig.2). If one cylindrical section will be damaged, the gas flows up (it is lighter then air), burns down only from this section (if film cannot easy burn) and piece get only hole. In any case the special margins do not allow the fire to set fire to next pieces.

(2) *Carbonic acid (smoke, $CO_2$) from industry and cars will pollute air into dome.*

The smoke from industry can be deleted out from dome by film tubes acting as feedthroughs (chimneys) to the outer air. The cars (exhaust pipes) can be provided by a carbonic acid absorber. The evergreen plants into Dome will intensely absorb $CO_2$ especially if concentration of $CO_2$ will be over the regular values in conventional atmosphere (but safe for people). We can also periodically ventilate the Dome in good weather by open the special windows in Dome (see fig.1) and turn on the ventilators like we ventilate the apartment. We can install heat exchangers and permanently change the air in the dome (periodically wise to do anyway because of trace contaminant buildups).

(3) *How can snow be removed from Dome cover?*

We can pump a warm air between the Blanket layers and melt show and pass the water by rain tubes. We can drop the snow by opening the Blanket windows (fig.2d).

(4) *How can dust be removed from the Dome cover?*

The Blanket is located at high altitude (about 500 m). Air at this altitude has but little dust. The dust that does infall and stick may be removed by rain, washdown tubes or air flow from blowers or even a helicopter close pass.

(5) *Storm wind overpressures?*

The storm wind can only be on the bounding (outside) sections of dome. Dome has special semi-spherical and semi-cylindrical form factor. We can increase the internal pressure in storm time to add robustness.

(6) *Cover damage.*

The envelope contains a rip-stop cable mesh so that the film cannot be damaged greatly. Electronic signals alert supervising personnel of any rupture problems. The needed part of cover may be reeled down by control cable and repaired. Dome has independent sections.

## Conclusion

The building of gigantic inflatable AB-Dome over an empty flat surface is not difficult. The cover spreads on said flat surface and a ventilator pumps air under the cover (the edges being joined and secured gas-tight) and the overpressure, over many hours, lifts the dome. However, if we want to

16cover a city, garden, forest we cannot easily spread the thin film over building or trees. In given article is suggested a new method which solves this problem. Idea is in design the double film Blanket filled by light gas (methane, hydrogen, helium). Subassemblies of the AB Dome, known as AB Blankets, are lighter then air and fly in atmosphere. They can be made on a flat area and delivered by dirigible or helicopter to the sky over the city. Here they are connected to the AB Dome under construction, cover the city and protect it from bad weather, chemical, biological and radioactive weapons and particulate falls. After finish of building the light gas can be changed by air.

*Acknowledgement*

The author wishes to acknowledge Joseph Friedlander (Israel) for correcting the author's English and offering useful technical advice.
# References

(The reader may find some of these articles at the author's web page: http://Bolonkin.narod.ru/p65.htm, http://arxiv.org , search term "Bolonkin", in the book "*Non-Rocket Space Launch and Flight*", Elsevier, London, 2006, 488 pgs., in book "*New Concepts, Ideas, Innovations in Aerospace, Technology and Human Science*", NOVA, 2007, 502 pgs., and in book "*Macro-Projects: Environment and Technology*", NOVA, 2008, 500 pgs.)

1. Bolonkin, A.A., (2003), "Optimal Inflatable Space Towers with 3-100 km Height", *Journal of the British Interplanetary Society* Vol. 56, pp. 87 - 97, 2003.
2. Bolonkin A.A., Cathcart R.B., (2006a), Inflatable 'Evergreen' Polar Zone Dome (EPZD) Settlements, 2006, http://arxiv.org search term is "Bolonkin".
3. Bolonkin, A.A., (2006b), Control of Regional and Global Weather, 2006, http://arxiv.org search for "Bolonkin".
4. Bolonkin A.A., (2006d), Cheap Textile Dam Protection of Seaport Cities against Hurricane Storm Surge Waves, Tsunamis, and Other Weather-Related Floods, 2006. http://arxiv.org.
5. Cathcart R.B. and Bolonkin, A.A., (2006e),. Ocean Terracing, 2006. http://arxiv.org.
6. Bolonkin A.A., (2006g), *Non-Rocket Space Launch and Flight, Elsevier*, London, 2006, 488 ps.
7. Bolonkin, A.A. and R.B. Cathcart, (2006i), Inflatable 'Evergreen' Dome Settlements for Earth's Polar Regions. Clean Technologies and Environmental Policy. DOI 10.1007/s10098.006-0073.4.
8. *Macro-Engineering: A Challenge for the Future*. Springer, (2006). 318 pages. Collection of articles. See articles of A. Bolonkin and R. Cathcart.
9. Bolonkin A.A., Cathcart R.B., (2007b), Inflatable 'Evergreen' dome settlements for Earth's Polar Regions. Journal "Clean Technologies and Environmental Policy", Vol 9, No. 2, May 2007, pp.125-132.
10. Bolonkin, A.A., (2007c),  "*New Concepts, Ideas, and Innovations in Aerospace, Technology and Human Life*". NOVA, 2007, 502 pgs.
11. Bolonkin A.A., (2007e), AB Method of Irrigation without Water (Closed-loop water cycle). Presented to http://arxiv.org in 2007 search "Bolonkin".
12. Bolonkin A.A., (2007f), Inflatable Dome for Moon, Mars, Asteroids and Satellites, Presented as paper AIAA-2007-6262 by AIAA Conference "Space-2007", 18-20 September 2007, Long Beach. CA, USA.
13. Bolonkin, A.A., (2007h), Cheap artificial AB-Mountains, Extraction of Water and Energy from Atmosphere and Change of Country Climate http://arxiv.org, 2007.
14. Bolonkin, A.A.,(2007i) "Optimal Inflatable Space Towers with 3-100 km Height", *Journal of the British Interplanetary Society* Vol. 56, pp. 87 - 97, 2003.
15. Bolonkin, A.A., Cathcart R.B., (2008b), *Macro-Projects: Environment and Technology*. NOVA, 500 pgs.
16. Bolonkin A.A., Cheap Method of City Protection from Rockets and Nuclear Warheads. http://arxiv.org search "Bolonkin". 2007.
17. Bolonkin A.A., Cheap Artificial AB-Mountains, Extraction of Water and Energy from Atmosphere and Change of Regional Climate.  Presented to http://arxiv.org in 2007 search "Bolonkin".
18. Gleick, Peter; et al. (1996). *Encyclopedia of Climate and Weather*. Oxford University Press.
19. Naschekin, V.V., (1969), *Technical thermodynamic and heat transmission*. Public House High University, Moscow. 1969 (in Russian).
17. *Wikipedia*. Some background material in this article is gathered from Wikipedia under the Creative Commons license.